\documentclass[%
	twocolumn,preprintnumbers,%
        aps,pra,showpacs,amsmath,amssymb]{revtex4}
\usepackage[pdftex]{graphicx}

\usepackage{amsmath,amssymb,amsfonts,amscd}
\usepackage[mathscr]{eucal}
\usepackage{bm}

\newcommand{\set}[1]{\left\{ #1 \right\}}

\newcommand{\bbset}[1]{\mathbb{#1}}
\newcommand{\Integer}{\bbset{Z}}

\newcommand{\Hilbert}{\mathcal{H}}

\def\toexp{\mathop{\rm exp}}
\newcommand{\Texp}{\toexp_{\leftarrow}}

\newcommand{\ket}[1]{|{}#1{}\rangle}
\newcommand{\bracket}[2]{\langle{}#1{}|{}#2{}\rangle}

\newcommand{\ketbra}[2]{|{}#1{}\rangle\langle{}#2{}|}

\newcommand{\Hq}{\Hilbert_{\rm b}}
\newcommand{\zero}{{\mathcal{Z}}}
\newcommand{\one}{{\mathcal{I}}}

\newcommand{\HilA}{\Hilbert_{\rm A}}
\newcommand{\HilC}{\Hilbert_{\rm C}}
\newcommand{\HilClock}{\Hilbert_{\rm clock}}

\newcommand{\HB}{\hat{H}_{\rm B}}
\newcommand{\HP}{\hat{H}_{\rm P}}
\newcommand{\EP}{E_{\rm P}}
\newcommand{\Ground}{0_{\rm B}}
\newcommand{\Answer}{x}

\newcommand{\Init}{0}
\newcommand{\Fin}{1}

\newcommand{\Fair}{\mathcal{F}}
\newcommand{\ctrlRegister}[1]{{#1}^{({\rm c})}}

\newcommand{\smax}{s_{\text{max}}}
\newcommand{\Tmax}{T_{\text{max}}}

\begin{document}

\title{Adiabatic quantum computation along quasienergies}
\author{Atushi Tanaka}
\email{tanaka-atushi@tmu.ac.jp}
\affiliation{Department of Physics, Tokyo Metropolitan University,
  Minami-Osawa, Hachioji, Tokyo 192-0397, Japan}

\author{Kae Nemoto}
\affiliation{National Institute of Informatics, 
  2-1-2 Hitotsubashi, Chiyoda-ku, Tokyo 101-8430, Japan}

\date{\today}

\begin{abstract}
  The parametric deformations of quasienergies and eigenvectors 
  of unitary operators are
  applied to the design of quantum adiabatic algorithms. 
  The conventional, {\em standard} adiabatic quantum computation proceeds
  along eigenenergies of parameter-dependent Hamiltonians.
  By contrast, {\em discrete adiabatic computation} utilizes
  adiabatic passage along the quasienergies of 
  parameter-dependent unitary operators.
  For example, such computation can be realized by 
  a concatenation of parameterized quantum circuits, with 
  an adiabatic though inevitably discrete change of the parameter.
  A design principle of adiabatic passage along quasienergy is
  recently proposed: Cheon's quasienergy and eigenspace anholonomies 
  on unitary operators is available to realize
  {\em anholonomic adiabatic algorithms}
  [Tanaka and Miyamoto, Phys. Rev. Lett. {\bf 98}, 160407 (2007)],
  which compose a nontrivial family of discrete adiabatic algorithms.
  It is straightforward to port a standard adiabatic algorithm
  to an anholonomic adiabatic one, except an
  introduction of a parameter $\ket{v}$, which is available 
  to adjust the gaps of the quasienergies to control 
  the running time steps.
  In Grover's database search problem, 
  the costs to prepare $\ket{v}$ for the qualitatively different,
  i.e., power or exponential, running time steps are shown to be 
  qualitatively different. 
  Curiously, in establishing the equivalence between
  the standard quantum computation based on the circuit model and
  the anholonomic adiabatic quantum computation model, it is shown that 
  the cost for $\ket{v}$ to enlarge the gaps of the eigenvalue is 
  qualitatively negligible.
\end{abstract}


\maketitle

\section{Introduction}
\label{sec:introduction}

An adiabatic passage along an eigenenergy of a Hamiltonian with 
slowly-varying parameter~\cite{AdiabaticPassage}
provides one of the simplest ways to control quantum states.
Recently, one of largest-scale applications of the adiabatic passage
to compose an algorithm for classically intractable problems is 
proposed by Farhi {\em et al.}~\cite{Farhi-quant-ph-0001106}.
Their adiabatic passage connects a quantum state, which is supposed 
to be easy to prepare, to a state that represents a solution of, 
for example, a satisfiability problem. 
Farhi {\em et al.} showed a systematic way to design a parameter-dependent 
Hamiltonian with local interactions of qubits only which governs the
adiabatic passage.
We will call this approach
{\em the standard} adiabatic quantum computation (SAQC)
as we later introduce a new approach to be distinguished from the SAQC.
Recently, the standard adiabatic quantum computation is proven to have
the same computational power as the standard quantum computation
in terms of
the computational complexity~\cite{Aharonov-SJOC-37-166}.
It is still an open question whether the adiabatic approach really 
solve the classically intractable problems efficiently 
as we can see the promising numerical results~%
\cite{Farhi-Science-292-472,Schuetzhold-PRA-74-060304}
as well as the reports of disastrous slowdown~%
\cite{Znidaric-PRA-71-062305,Znidaric-PRA-73-022329,Farhi-IJQI-6-503}.

In many studies of SAQC, 
e.g., 
in the simulations of SAQC 
by quantum circuits~\cite{Farhi-quant-ph-0001106}
and classical digital computers~\cite{Hogg-PRA-67-022314},
or
in experimental realizations where the adiabatic changes
of coupling constants are 
infeasible~\cite{Steffen-PRL-90-067903,note:discretizeSAQC},
it is inevitable to introduce 
{\em the discretization of the adiabatic deformation} of parameters.
In such cases, the time evolution in the computational process is
realized by the products of parameterized unitary operators, each of
which represents a computational step to emulate 
the time evolution for the unit of time.
Due to the tolerance of the adiabatic passage against small
perturbations, the discretization of the time evolution, in general,
provides a good approximation of SAQC.  
The discretization process
however can be considered to introduce an alternative model of
adiabatic quantum computation.
The state vector follows the adiabatic change of an eigenvector
of a unitary operator, where the definition of eigenenergy can be 
inapplicable even in any approximate sense.
This scheme, which we call discrete adiabatic quantum computation
(DAQC), will be formulated based on the adiabatic passage 
along a quasienergy~\cite{Zeldovich-JETP-24-1006},
or equivalently, an eigenangle of slowly-varying unitary operator.

DAQC is useful to generalize quantum adiabatic algorithm, 
which we will show in this paper.
A family of DAQC that essentially relies on
the adiabatic passage along a quasienergy is recently proposed by 
one of the authors~\cite{Tanaka-PRL-98-160407}. 
Here the adiabatic passage is composed with a help of
Cheon's eigenvalue anholonomy%
~\cite{Cheon-PLA-248-285,Tanaka-PRL-98-160407,Miyamoto-PRA-76-042115,%
  Cheon-EPL-85-20001},
which enables us to design adiabatic passages that visit
all eigenstates of a given unperturbed Hamiltonian.
This adiabatic scheme, which will be called 
anholonomic adiabatic quantum computation (AAQC),
composes an interesting and nontrivial family of DAQC.
In particular, AAQC does not approximate SAQC, and
hence the question whether AAQC is equivalent to
the standard quantum computation naturally arises~\cite{Tanaka-PRL-98-160407}. 

In this paper, we establish a formulation of AAQC and elucidate 
its equivalence to the standard quantum computation.
To achieve this, we start from the formulation of DAQC, 
the most general family of
the adiabatic algorithm, in Section~\ref{sec:DAQC}.
It turns out that AAQC, which composes a subset of DAQC,
offers us a systematic way to design 
nontrivial instances of DAQC (Section~\ref{sec:FundamentalsAAQC}). 
We show how the performance of AAQC can be evaluated
in the Grover's unstructured database search problem%
~\cite{Grover-PRL-79-325}. 
It is shown that an ingredient of AAQC strongly affects the performance
that is determined by the ``gap'' of eigenvalues
(Section~\ref{sec:AAQCExamples}).
In Section~\ref{sec:AAQCEquivalence}, we show that
the anholonomic adiabatic quantum computation model is
equivalent with the standard quantum computation, through a modification 
of Aharonov {\em et al.}'s proof~\cite{Aharonov-SJOC-37-166} of 
the equivalence between the standard quantum computation and
the standard adiabatic computation model.
There, it turns out that the control of the gap in AAQC
discussed in Section~\ref{sec:AAQCExamples} plays 
crucial role to show the equivalence.

\section{Discrete Adiabatic Quantum Computation}
\label{sec:DAQC}

In this section, we will introduce 
discrete adiabatic quantum computation (DAQC) 
in order to facilitate
our study of AAQC in the following sections.
First, a formulation of DAQC, which offers a unified framework for 
the adiabatic algorithm, is established 
in Section~\ref{subsec:DAQCformulation}.
Second, it is shown that a family of DAQC approximates SAQC 
in Section~\ref{subsec:DAQCfromSAQC}.
Third, we show the equivalence of DAQC and the standard quantum 
computation in Section~\ref{subsec:DAQCequivalence}.

\subsection{Formulation of DAQC}
\label{subsec:DAQCformulation}

We will design a computational system that involves $n$ qubits. 
This naturally 
introduces the concept of locality among qubits in the computational system.
Namely, if an operation involves only few, $n$-independent numbers of qubits,
the operation is called {\em local}.
The time evolution of the quantum state is
governed by a unitary operator $\hat{U}_{s}$ 
with an ``adiabatic'' parameter $s$.
In the stroboscopic description of the time evolution,
a unit step evolution of a quantum state from $\ket{\psi'}$ to
$\ket{\psi''}$ is described
by a quantum map~\cite{Berry-AP-122-26}
\begin{equation}
  \label{eq:quantumMap}
  \ket{\psi''} = \hat{U}_{s}\ket{\psi'}  
  .
\end{equation}
Since the quantum map is iterated extensively 
to realize an adiabatic passage,
$\hat{U}_{s}$ is required to be {\em efficiently} implementable,
whose definition will be introduced so as to be compatible with the one
for quantum circuits.
First, we explain a construction of $\hat{U}_{s}$ from
a product of local unitary operations.
The unitary operator $\hat{U}_{s}$ accordingly has a natural 
counterpart of a quantum circuit,
where each factor of $\hat{U}_{s}$ corresponds to an element of 
the circuit.
If the number of 
the factors is bounded asymptotically by a polynomial of $n$, 
$\hat{U}_{s}$ is called to be {\em efficiently} implementable.
The number of the 
iterations of the quantum map characterizes the computational 
complexity of the algorithm.

Next we discuss 
the case that
$\hat{U}_{s}$ is realized 
by a time evolution induced by a time-dependent Hamiltonian 
$\hat{H}_{s}(t)$ during a finite time interval (say, $0\le t \le T$),
i.e.,
\begin{equation}
  \hat{U}_{s} = 
  \Texp\left(-\frac{i}{\hbar}\int_0^{T}\hat{H}_{s}(t) dt\right),
\end{equation}
where $\Texp$ denotes the time-orderd exponential.
We restrict ourselves to the case that 
$\hat{H}_{s}(t)$ is composed by 
a finite number of local interaction terms. 
In particular, we call $\hat{H}_{s}(t)$ efficiently implementable,
when the number of the local interaction terms for $\hat{H}_{s}(t)$
is bounded by a polynomial of $n$. 
Based on the argument 
by Lloyd~\cite{Lloyd-Science-273-1073,Farhi-quant-ph-0001106},
the efficiency discussed here is equivalent to the efficiency for 
quantum circuits mentioned above.
In addition to the two approaches of the implementation of $\hat{U}_s$,
it is possible to merge two approaches to construct $\hat{U}_s$.
An example will be shown in Section~\ref{sec:AAQCEquivalence}.

To introduce an adiabatic passage for a quantum map, the eigenvalue problem
of $\hat{U}_{s}$ is explained more precisely.
An eigenvalue of $\hat{U}_{s}$ is 
a unimodular complex number, due to the unitarity. 
We parameterize the eigenvalue $e^{-i\theta(s)}$ 
by a real number $\theta(s)$,
which is called an eigenangle~\cite{note:eigenangle}.
By definition, the space of eigenangle has a period of $2\pi$.
Since we assumed in the following that
a unit time interval $T$ is associated with the quantum
map $\hat{U}_{s}$, e.g. $\hat{U}_{s}$ is induced
by a Hamiltonian time evolution during time interval $T$, 
an eigenangle $\theta(s)$ provides a 
characteristic energy scale $E(s)\equiv\theta(s)\hbar T^{-1}$, which is 
called quasienergy~\cite{Zeldovich-JETP-24-1006} and determined
up to modulo $2\pi\hbar{}T^{-1}$.
Corresponding to $E(s)$,
let $\ket{\xi(s)}$ be an eigenvector of $\hat{U}_{s}$,
i.e., they satisfy the eigenvalue equation
$\hat{U}_{s}\ket{\xi(s)}=
e^{-i T E(s)/\hbar}\ket{\xi(s)}$.
We further require that $E(s)$ and $\ket{\xi(s)}$
are smooth with respect to $s$.
The eigenvector $\ket{\xi(s)}$ is also called a stationary state,
due to its characteristic in dynamics.
For the fixed value of $s$, the state $\ket{\xi(s)}$ is invariant under 
the time evolution induced by successive application of $\hat{U}_{s}$
up to the accumulated dynamical phase. 

To explain a discrete adiabatic behavior of the state vector,
let us change the value of $s$ from $0$ to $s_{\rm max}$, 
under a successive application of $\hat{U}_{s}$
during $L$ ($\gg 1$) steps.
Let $s_l$ be the value of $s$ at $l$-th step, where
$s_0 = 0$ and $s_L=s_{\rm max}$.
The exact final state is
\begin{equation}
  \ket{\Psi_L}\equiv
  \hat{U}_{s_{L}}\hat{U}_{s_{L-1}}
  \ldots\hat{U}_{s_{1}}\ket{\xi(0)}.
\end{equation}
The adiabatic theorem for quantum maps~\cite{Hogg-PRA-67-022314}
ensures that
$\ket{\Psi_L}$ converges to $\ket{\xi(s_{\rm max})}$ up to the global phase 
as $L\to \infty$, if the gaps between the nearest eigenangles are 
large enough.
In reality, $L$ is always kept to be finite and the adiabatic time
evolution is consequently not exact. 
A criterion to determine $L$ is the condition
that the time evolution obeys the adiabatic time evolution within 
an error $\epsilon$,
i.e., 
$\bm{\|}\ketbra{\Psi_L}{\Psi_L}
-\ketbra{\xi(s_{\rm max})}{\xi(s_{\rm max})}\bm{\|}
< \epsilon$. 
This is equivalent with $L \ge L(\epsilon)$, where
$L(\epsilon)$ is an appropriate lower bound.
Here, the most important source of the error in the asymptotic regime
$L\to\infty$ is nonadiabatic transition~\cite{Nonadiabatic}. 
Hence the crudest estimate of $L(\epsilon)$ will be determined 
by the Landau-Zener-St\"uckelberg formula
with a modification that comes from 
the use of quasienergies~\cite{Breuer-ZPD-11-1},
instead of eigenenergy,
for the discrete adiabatic processes.

To summarize the above argument, a definition of discrete adiabatic 
computation (DAQC) is given:
An instance of DAQC is specified by an efficiently 
implementable unitary operator $\hat{U}_{s}$, an initial 
state $\ket{\Psi_0}$, and a schedule $\set{s_l}_{l=0}^L$
of the adiabatic parameter from $s_0 = 0$ to $s_L = s_{\rm max}$
with the number of steps $L$. 
DAQC is designed to transport the initial state vector 
$\ket{\Psi_0}$, which is the eigenstate $\ket{\xi(s=0)}$ of 
$\hat{U}_0$, to the target state $\ket{\xi(s=s_{max})}$ within
an error $\epsilon$ through the continuous deformation of an 
eigenstate $\ket{\xi(s)}$ of $\hat{U}_s$.
The initial state is be prepared from $\ket{0}$ by an efficiently 
implementable unitary operator $\hat{U}_{\rm I}$, 
i.e. $\ket{\Psi_0}=\hat{U}_{\rm I}\ket{0}$.
We need to make $L$ large enough so that 
the final state 
$\ket{\Psi_L}=\hat{U}_{s_{L}}\hat{U}_{s_{L-1}}
\ldots\hat{U}_{s_{1}}\ket{\Psi_0}$ 
of the computation
is a good approximation of 
the target state of the computation $\ket{\xi(s_{\rm max})}$.
The running time $L(\epsilon)$ is the minimum value of 
$L$ to achieve that the distance between the final state 
$\ket{\Psi_L}$ and $\ket{\xi(s_{\rm max})}$ is smaller than 
a given error $\epsilon$.

\subsection{DAQC from SAQC through 
  the discretization of continuous-time evolution}
\label{subsec:DAQCfromSAQC}

As a way to construct DAQC,
the discretizations of standard adiabatic 
quantum computation (SAQC) are explained for a pedagogic purpose.
As mentioned in the introduction, many studies of SAQC are 
done through the
discretization of the variation of the adiabatic parameter.
The discretized time evolution is introduced so that it
approximates well the ideal, continuous-time evolution.
We first explain the standard adiabatic quantum
computation.
A computational system is specified by a Hamiltonian $\hat{H}(s)$, 
which can be decomposed into a finite number of local Hamiltonians, 
with an adiabatic parameter $s$. 
We assume, without loss of generality, that the initial and the final
values of $s$ are $0$ and $\smax$, respectively.
The initial state $\ket{\Psi_0}$ must be the ground state of 
$\HB\equiv\hat{H}(0)$,
which can be efficiently prepared.
The target of the computation is the ground state of
$\HP\equiv\hat{H}(\smax)$. 
For a given time-dependence of $s(t)$ with $0\le t\le \Tmax$,
$s(0)=0$, and $s(\Tmax)=\smax$,
the exact final state of the computation (of the unitary evolution
part) is
\begin{equation}
  \label{eq:SAQCevolv}
  \Texp\left(-i\int_0^{\Tmax} \hat{H}\bm{(}s(t)\bm{)} dt\right)\ket{\Psi_0}.
\end{equation}
In the adiabatic limit $\Tmax\to\infty$, the finial state converges to
the target for the final state.
The running time of the computation is the minimum value of $\Tmax$
that keeps the distance between the finial and the target states
within a given error $\epsilon$.

A discretization of the time-dependence in $s(t)$ 
gives us an instance of DAQC. For a discretization of time
$t_l$ ($0\le l \le L$), we introduce a piecewise constant
function $\tilde{s}(t) = s_l \equiv \left[s(t_{l-1})+s(t_{l})\right]/2$
for $t_{t-1} < t < t_l$. For the sake of simplicity,
we set $s_0\equiv 0$ and $t_{-1}\equiv -t_1$. Accordingly, we have
a unitary operator of a time interval $(t_{l-1},t_{l})$:
\begin{equation}
  \hat{U}_l \equiv \exp\left[-i \hat{H}(s_l)(t_{l} - t_{l-1})\right].
\end{equation}
The final state of a SAQC with the schedule $\tilde{s}(t)$ is
\begin{equation}
  \hat{U}_L\hat{U}_{L-1}\ldots\hat{U}_{1}\ket{\Psi_0}.
\end{equation}
Namely, we obtain a DAQC from a SAQC, since $\ket{\Psi_0}$ is
also an eigenvector of $\hat{U}_0 = e^{-i\hat{H}(0) (t_0 - t_{-1})}$.
We may expect the convergence of a DAQC to the original
SAQC in the limit $L\to\infty$. For finite $L$, the DAQC 
is an approximation of the SAQC. Hence we may regard that the DAQC
approximately follows the adiabatic passage built on the
parametrically varying eigenenergy.

\subsection{DAQC is equivalent to
  the standard quantum computation}
\label{subsec:DAQCequivalence}

Now we examine the equivalence of DAQC and the standard quantum computation,
in the sense of computational complexity.
First, it is straightforward to see that 
a DAQC is efficiently simulated by a quantum circuit
from the definition above.
Second, for the inverse argument, i.e., whether a quantum circuit can be 
simulated by a DAQC, we need to remind that DAQC involves a rather wide 
class of computation. In particular, as is explained above,
DAQC contains the discretizations of the standard adiabatic
quantum computation, which are proven to be equivalent with
the standard quantum computation~\cite{Aharonov-SJOC-37-166}. 
Hence, it is trivial to conclude that 
DAQC are equivalent with the standard quantum computation.

Much more serious question on the equivalence arises when we
examine a subclass of DAQC, by introducing an alternative design
of DAQC, where the adiabatic passage built on the parametrically varying 
{\em eigenenergy} is inapplicable.
Instead, an adiabatic passage built on a {\em quasienergy} needs to 
be employed. 
We will discuss such an example in the next section.

\section{Anholonomic adiabatic quantum computation}
\label{sec:FundamentalsAAQC}

In this section, anholonomic adiabatic quantum computation (AAQC), 
which compose a subclass of DAQC, is introduced.
The adiabatic passages of AAQC can be constructed 
with the help of Cheon's eigenvalue and eigenspace 
anholonomies, which will be explained 
in Section~\ref{subsec:QuasienergyAnholonomy}.
Subsequently, a formulation of AAQC is introduced 
in Section~\ref{subsec:FormulationOfAAQC}.
Finally, the simulation of AAQC by quantum circuits is examined
in Section~\ref{subsec:SimulationOfAAQCbyCircuits}, where a crucial
ingredient to determine the cost of the simulation is identified.

\subsection{Cheon's anholonomies for unitary operators}
\label{subsec:QuasienergyAnholonomy}

First of all, we explain Cheon's eigenvalue and eigenspace anholonomies%
~\cite{Cheon-PLA-248-285} and their realization in
a periodically pulsed system under a rank-$1$ 
perturbation~\cite{Tanaka-PRL-98-160407}.
This
offers a systematic 
design principle for adiabatic passages, in particular, adiabatic quantum
computation, along parametric changes of unitary 
operators~\cite{Miyamoto-PRA-76-042115}.
We assume that the spectrum of an ``unperturbed'' Hamiltonian $\hat{H}_0$
is discrete, finite and non-degenerate.
We will examine a periodically-pulsed driven system
described by the following ``kicked'' Hamiltonian:
\begin{equation}
  \label{eq:kickHamiltonian}
  \hat{H}_{s}(t) \equiv \hat{H}_0 
  + s\ketbra{v}{v}\sum_{n\in\Integer}\delta(t -n T),
\end{equation}
where $T$ and $s$ are the period and the strength of
the perturbation, and $\ket{v}$ is assumed to be normalized.
The kicked Hamiltonian (\ref{eq:kickHamiltonian}) naturally induces
a quantum map $\ket{\psi_{n+1}} = \hat{U}_{s}\ket{\psi_{n}}$,
where $\ket{\psi_n}$ is the state
vector just before the kick at $t = nT$ and 
\begin{equation}
  \label{eq:rank1Floquet}
  \hat{U}_{s}\equiv e^{-i\hat{H}_0 T}e^{-is\ketbra{v}{v}}
\end{equation}
is the corresponding Floquet operator.
We set $\hbar = 1$ in the following.

We remark the topological structure of the parameter space $s$
for the Floquet operator $\hat{U}_{s}$~(\ref{eq:rank1Floquet}).
From
$e^{-is\ketbra{v}{v}} = (1-\ketbra{v}{v}) + e^{-is}\ketbra{v}{v}$
,
both $\hat{U}_{s}$ and its spectrum
is periodic in $s$ with period $2\pi$.
Hence we may identify the parameter space of $s$ as a circle.

We explain what
will happen for each quasienergy and each eigenvector of
$\hat{U}_{s}$~(\ref{eq:rank1Floquet})
when we increase $s$ for a period.
Let us consider the ground energy $E'$ 
and the first excited eigenenergy
$E''$ of $\hat{H}_0$ ($E' < E''$)~\cite{note:twoquasienergy}. 
The corresponding normalized eigenvectors are denoted
by $\ket{E'}$ and $\ket{E''}$, respectively.
For simplicity, we employ a normalized vector
$\ket{v} = a\ket{E'}+b\ket{E''}$, where $a\ne0$, $b\ne0$ and $|a|^2+|b|^2=1$,
so that 
the subspace spanned by
$\ket{E'}$ and  $\ket{E''}$ is
invariant under $\hat{U}_{s}$. 
Accordingly, in the subspace, which we will focus on,
$\hat{U}_{s}$ is a $2$-dimensional matrix effectively.
Let $\ket{\xi_0(s)}$ be the eigenvector of $\hat{U}_{s}$
with $\ket{\xi_0(0)} = \ket{E'}$ and $E_0(s)$ the corresponding
quasienergy, whose branch is chosen as $[E', E'+2\pi T^{-1})$,
i.e., $E' < E'' < E+2\pi T^{-1}$.
Since $\partial_{s} E_0(s) =
T^{-1} \bracket{\xi_0(s)}{v}\bracket{v}{\xi_0(s)}$ 
is positive, $E_0(s)$ increases as $s$ increases.
At the same time, 
because of the periodicity of $\hat{U}_{2\pi} = \hat{U}_{0}$,
$E_0(2\pi) = E'+ 
\int_{0}^{2\pi} \left[\partial_{s} E_0(s)\right]ds$
must be equal to $E'$ or $E''$ (modulus $2\pi T^{-1}$).
The above choice of $\ket{v}$ ensures 
$0 < \partial_{s} E_0(s) < 2\pi T^{-1}$,
i.e., $E' < E_0(2\pi) < E'+2\pi T^{-1}$.
Hence we have $E_0(2\pi) = E''$, which implies a presence of
quasienergy anholonomy. The corresponding eigenvector 
$\ket{\xi_0(s)}$ arrives at $\ket{E''}$ at $s = 2\pi$.
Thus the minimum example of a path along quasienergy anholonomy
$E_0(s)$ is shown. Along the path, an adiabatic increment
of $s$ for a period of its parameter space
(i.e., from $0$ to $2\pi$) transfers the state vector that 
is initially prepared to be $\ket{E'}$, to $\ket{E''}$.

So far, we did not specify $\ket{v}$, except that it is a superposed
state of $\ket{E'}$ and $\ket{E''}$, because the quasienergy anholonomy
do not depend on the details of $\ket{v}$.
For example, the quasienergy anholonomy persists even under adiabatically 
slow fluctuations of $\ket{v}$. On the other hand, $\ket{v}$ 
affects
on the precise shape of $E_0(s)$ and adjacent quasienergy curves.
Namely, $\ket{v}$ determines the quasienergy gaps between neighboring 
levels of $E_0(s)$ and accordingly a timescale for adiabatic
passages.

Extensions of the quasienergy anholonomy for multilevels are 
straightforward~\cite{Tanaka-PRL-98-160407,Miyamoto-PRA-76-042115}.
For example, if we choose $\ket{v}$ that has non-zero overlapping 
with all eigenvectors of $\hat{H}_0$, 
$E_0(2\pi n)$ is the $n$-th excited energy of 
$\hat{H}_0$~\cite{note:nanholonomy}.
Thus $E_0(s)$ connects all the eigenvalues of $\hat{H}_0$
to offer a path for adiabatic passages with adiabatic increments of $s$.

\subsection{Formulation of AAQC}
\label{subsec:FormulationOfAAQC}

An anholonomic adiabatic quantum algorithm 
can be described based on the standard adiabatic algorithm. 
The standard adiabatic algorithm is specified by
a parameter-dependent Hamiltonian $H(s)$
on the Hilbert space $\HilA$.
Here, we employ the
simplest case that $H(s)$ is linear in $s$:
$\hat{H}(s) = (1-s)\hat{H}_{\rm B} + s \hat{H}_{\rm P}$,
where $\HB$ and $\HP$ are
the initial and the final Hamiltonians, respectively.
We impose that $\HB$ has non-degenerate ground 
state $\ket{\Ground}$ with the ground energy $0$.
The finial Hamiltonian $\HP$ is a cost function
of the problem to solve. Namely, an eigenvalue of $\HP$
indicates a ``distance'' between the corresponding eigenvector and
the answer of the problem.
We assume that the answer is unique and let $\ket\Answer$ be the
ground state of $\HP$.
To reuse $\HB$ and $\HP$ in the anholonomic approach, 
the maximum eigenvalues of $\HB$ and $\HP$ need to be
finite.

We now explain an implementation of an anholonomic adiabatic quantum 
processor under the assumption that the standard adiabatic quantum 
processor is available.
The state space $\HilA$, and its Hamiltonians
$\hat{H}_{\rm B}$ and  $\hat{H}_{\rm P}$
are reused. An additional qubit is employed as a ``control register,''
whose Hilbert space $\HilC$ has a complete orthonormal system
$\set{\ket\Init, \ket\Fin}$. 
Hence the whole Hilbert space is $\HilA\otimes\HilC$. 
Corresponding to the ``initial'' and ``final'' states of computation,
we introduce two projectors
$\ctrlRegister{\hat\zero}\equiv\ketbra\Init\Init$
and $\ctrlRegister{\hat\one}\equiv\ketbra\Fin\Fin$, respectively,
on $\HilC$.
Combining these parts, we obtain
an unperturbed Hamiltonian
\begin{equation}
  \label{eq:unperturbedAAC}
  \hat{H}_0 \equiv 
  \HB\otimes\ctrlRegister{\hat\zero}
  + (E_{\rm P} + \HP)\otimes\ctrlRegister{\hat\one},
\end{equation}
where we assume that $E_{\rm P}$ is positive and smaller than
the first excited energy of $\HB$.
It is straightforward to see that 
the ground energy $0$ and the first excited energy $E_{\rm P}$
of $\hat{H}_0$ are nondegenerate.
By construction, the ground and the first excited states of $\hat{H}_0$ 
are 
$\ket{-}\equiv \ket\Ground\otimes\ket\Init$
and $\ket{+}\equiv \ket\Answer\otimes\ket\Fin$, respectively.
We employ the quantum map~(\ref{eq:rank1Floquet}) to realize
the quasienergy anholonomy that connects between
the ground energy $0$ and the first excited energy $E_{\rm P}$
of $\hat{H}_0$~(\ref{eq:unperturbedAAC}). 
We impose the conditions $\bracket{\pm}{v}\ne 0$ for $\ket{v}$
so that the quasienergy $E_0(s)$ of $\hat{U}_{s}$
with $E_0(0) = 0$ reaches $E_{\rm P}$ at $s=2\pi$.
The adiabatic passage along $E_0(s)$ ($0\le{}s\le2\pi$)
transfers the state vector from $\ket{-}$ to $\ket{+}$, 
then the final state
provides the answer.
To evaluate the performance of the anholonomic adiabatic quantum
computation, we need to obtain the quasienergy gap with respect 
to the path $E_0(s)$. 
The quasienergy gap depends on the choice of $\ket{v}$.
We will evaluate the gaps for two choices of $\ket{v}$
in the following sections.

\subsection{Costs for simulations of AAQC by quantum circuits}
\label{subsec:SimulationOfAAQCbyCircuits}

We show how the unitary operator 
$\hat{U}_{s}$~\eqref{eq:rank1Floquet} of
the anholonomic adiabatic quantum processor is
composed by local quantum circuits, in a similar way explained
in Farhi {\em et al.}~\cite{Farhi-quant-ph-0001106}.
From the definition of $\hat{H}_0$~\eqref{eq:unperturbedAAC},
$\hat{U}_{s}$ is expressed by a product
$\hat{U}_{s}
= \hat{U}_{0 {\rm B}}\; \hat{U}_{0 {\rm P}}
\; e^{-is\ketbra{v}{v}}$,
where 
$\hat{U}_{0 {\rm B}}\equiv 
e^{-i\hat{H}_{\rm B} \otimes \ctrlRegister{\hat\zero} T}$
and 
$\hat{U}_{0 {\rm P}}\equiv 
e^{-i (\hat{H}_{\rm P} + E_{\rm P}) \otimes \ctrlRegister{\hat\one} T}$.

In the analysis of $\hat{U}_{0 {\rm B}}$, 
we assume that $\hat{H}_{\rm B}$ is composed by
$\hat{H}_{{\rm B},j}$, initial Hamiltonians for $j$-th qubit.
For example, it is often the case 
that $\hat{H}_{{\rm B},j}$ is an Hadamard operation.
This allows further decomposition:
\begin{equation}
  \hat{U}_{0 {\rm B}}
  = 
  e^{-i\hat{H}_{{\rm B},0} \otimes \ctrlRegister{\hat\zero} T}
  e^{-i\hat{H}_{{\rm B},1} \otimes \ctrlRegister{\hat\zero} T}
  \cdots
  e^{-i\hat{H}_{{\rm B},{n-1}} \otimes \ctrlRegister{\hat\zero} T},
\end{equation}
where 
$e^{-i\hat{H}_{{\rm B},j} \otimes \ctrlRegister{\hat\zero} T}
= (1 - \ctrlRegister{\hat\zero}) 
+ \ctrlRegister{\hat\zero} e^{-i\hat{H}_{{\rm B},j} T}$
is a controlled-$1$ bit unitary operation.
Hence $\hat{U}_{0 {\rm B}}$ is simulated by $n$ controlled-$1$ bit 
unitary gates.

To examine $\hat{U}_{0 {\rm P}}$, 
assume that $\HP$ is a cost Hamiltonian of a satisfiability problem, 
which is composed by clauses 
$C_j$. 
Let $m$ be the number of the clauses
and $\hat{H}_{{\rm P},j}$ a corresponding local 
Hamiltonian for $C_j$. 
Note that 
$\hat{H}_{{\rm P}}=\sum_j\hat{H}_{{\rm P},j}$,
and the elements $\hat{H}_{{\rm P},j}$
commute with each other.
Hence we have a decomposition of $\hat{U}_{0 {\rm P}}$:
\begin{align}
  \hat{U}_{0 {\rm P}}&
  =
  e^{-i\EP \ctrlRegister{\hat\one} T}
  e^{-i\hat{H}_{{\rm P},0} \otimes \ctrlRegister{\hat\one} T}
  \nonumber \\ &{}\qquad\times
  e^{-i\hat{H}_{{\rm P},1} \otimes \ctrlRegister{\hat\one} T}
  \cdots
  e^{-i\hat{H}_{{\rm P},m-1} \otimes \ctrlRegister{\hat\one} T}
\end{align}
where
$e^{-i\hat{H}_{{\rm P},j} \otimes \ctrlRegister{\hat\one} T}
= (1 -\ctrlRegister{\hat\one}) 
+ \ctrlRegister{\hat\one}e^{-i\hat{H}_{{\rm P},j}}$
is 
a several qubits
unitary operation with a condition
(3 qubit unitary operation for 3-SAT problem).
Hence, a simulation of $\hat{U}_{0 {\rm P}}$ requires
a conditional phase-shift gate and
$m$ conditional, few bits unitary gates.

Finally, $e^{-is\ketbra{v}{v}}$ is examined.
With two unit vectors $\ket{v_{\Init}}$ and $\ket{v_{\Fin}}$ in $\HilA$,
$\ket{v}$ is written as
$
  \ket{v} = 
   c_{\Init} \ket{v_{\Init}}\otimes\ket\Init
  +c_{\Fin}\ket{v_{\Fin}}\otimes\ket\Fin  
$,
where $c_{\Init}$ and $c_{\Fin} $ are coefficients.
We
introduce two unitary operators $\hat{G}_{\Init}$ and $\hat{G}_{\Fin}$
so as to induce  $\ket{v_{\Init}}$ and $\ket{v_{\Fin}}$
from $\ket{0}\in\HilA$, i.e.,
$\ket{v_{\Init}} = \hat{G}_{\Init}\ket{0}$ and 
$\ket{v_{\Fin}} = \hat{G}_{\Fin}\ket{0}$.
Hence we have $\ket{v} =  \hat{G}\left(\ket{0}\otimes\ket{\rm C}\right)$,
where $\hat{G}\equiv 
\left(\hat{G}_{\Init}\otimes\ctrlRegister{\hat\zero}+\ctrlRegister{\hat\one}
\right)
\left(\ctrlRegister{\hat\zero}+\hat{G}_{\Fin}\otimes\ctrlRegister{\hat\one}
\right)$
and 
$\ket{\rm C}\equiv c_{\Init}\ket\Init  + c_{\Fin}\ket\Fin$,
and
\begin{align}
  e^{-is\ketbra{v}{v}}
  = \hat{G} 
  e^{-is\ketbra{0}{0}\otimes\ketbra{\rm C}{\rm C}} \hat{G}^{\dagger},
\end{align}
where $e^{-is\ketbra{0}{0}\otimes\ketbra{\rm C}{\rm C}}$
is a phase-shift gate.
Note that $\hat{G}$ is a product of two conditional-unitary operations.
If both $\hat{G}_{\Init}$ and  $\hat{G}_{\Fin}$ are 
simulated efficiently by quantum circuits, in other words,
both $\ket{v_{\Init}}$ and $\ket{v_{\Fin}}$ 
can be 
efficiently
prepared
by the standard quantum computers,
$\hat{G}$ 
can 
also 
be
simulated efficiently.
For example, 
a state $2^{-n/2}\sum_{j=0}^{2^n-1}\ket{j}$ in $\HilA$ is
efficiently available, since the state is $H^{\otimes n}\ket{0}$
and is available through $n$ Hadamard operations $H$.
We may employ $H^{\otimes n}\ket{0}$
for $\ket{v_{\Init}}$, and, in Section~\ref{sec:smallgap},
we also employ $H^{\otimes n}\ket{0}$ for $\ket{v_{\Fin}}$.
On the other hand, in Section~\ref{sec:widegap}, 
we employ $\ket{v_{\Fin}}= \ket{x}$, which is the answer of the problem.
Note that if $\ket{x}$ is efficiently available with quantum circuits, 
$\hat{G}$ is also efficiently available with quantum circuits
and we may put $\hat{G}$ in our anholonomic processor only with
a qualitatively negligible cost.
Otherwise, the implementation of $\hat{G}$ requires 
exponentially many quantum gates
by the definition of the efficiency here.

\section{Two examples of AAQC}
\label{sec:AAQCExamples}
Through the study of two examples,
we show that the magnitude of the minimum quasienergy gap drastically
depends on $\ket{v}$. At the same time, we examine the cost
for the increment of the gap, in terms of the number of
quantum gates to realize $\ket{v}$.

\subsection{An ``optimal'' choice of $\ket{v}$ to widen 
  the quasienergy gap}
\label{sec:widegap}

Let us assume that the unique solution of a given problem $\ket{x}$
is available to construct 
$\ket{v}$ for the anholonomic processor:
\begin{align}
  \label{eq:impracticableV}
  \ket{v}&
  \equiv
  \frac{1}{\sqrt{2}}
  \left(\ket\Ground\otimes\ket{\Init}+\ket\Answer\otimes\ket{\Fin}\right)
  = 
  \frac{\ket{-}+\ket{+}}{\sqrt{2}}
  .
\end{align}
As is seen below, the point of the assumption is that $\ket{v}$ is
a superposed state of $\ket{\pm}$, and the precise values of the coefficients
are irrelevant.
The following argument is rather general 
since it is applicable as long as the problem has a unique solution.

First, we examine the quasienergy gap during the adiabatic passage.
With our choice of $\ket{v}$~\eqref{eq:impracticableV},
the linear span of $\ket{\pm}$ is invariant under
$\hat{U}_{s}$~(Eq.~\eqref{eq:rank1Floquet}), 
since $\ket{\pm}$ are eigenstates of the unperturbed part $\hat{U}_{0}$ 
and the linear span of $\ket{\pm}$ is invariant under
the kick part $e^{-is\ketbra{v}{v}}$ of $\hat{U}_{s}$,
due to the fact
\begin{align}
  e^{-is\ketbra{v}{v}}&
  = \frac{1}{2}\left(1 -i\hat\sigma_y\right)
  + e^{-is}\frac{1}{2}\left(1 + \hat\sigma_x\right)
  ,
\end{align}
where
$\hat\sigma_x\equiv\ketbra{-}{+} + \text{h.c.}$
and $\hat\sigma_y\equiv i\ketbra{+}{-} + \text{h.c.}$.
Hence it is suffice to examine $\hat{U}_{s}$ within the subspace spanned by
$\ket{\pm}$.
The characteristic equation of $\hat{U}_{s}$ within the subspace is
\begin{align}
  &
  \det\left| z -
    \begin{bmatrix}
      1& 0 \\ 0& e^{-i\EP T}
    \end{bmatrix}
    \times e^{-is/2}
    \begin{bmatrix}
      \cos\frac{s}{2}& -i\sin\frac{s}{2}\\
      -i\sin\frac{s}{2}& \cos\frac{s}{2}
    \end{bmatrix}
  \right|
  \nonumber \\ &
  = 0
  .
\end{align}
The above characteristic equation is independent of the details of
$\HP$, in particular, the size of the problem $n$.
This indicates that the quasienergy gap is also independent of $n$.
Hence,
the choice of $\ket{v}$
above~(\ref{eq:impracticableV}) provides an extraordinary speed up 
of the anholonomic processor for asymptotically large $n$.

\begin{figure}
  \centering
  \includegraphics[width=8.6cm]{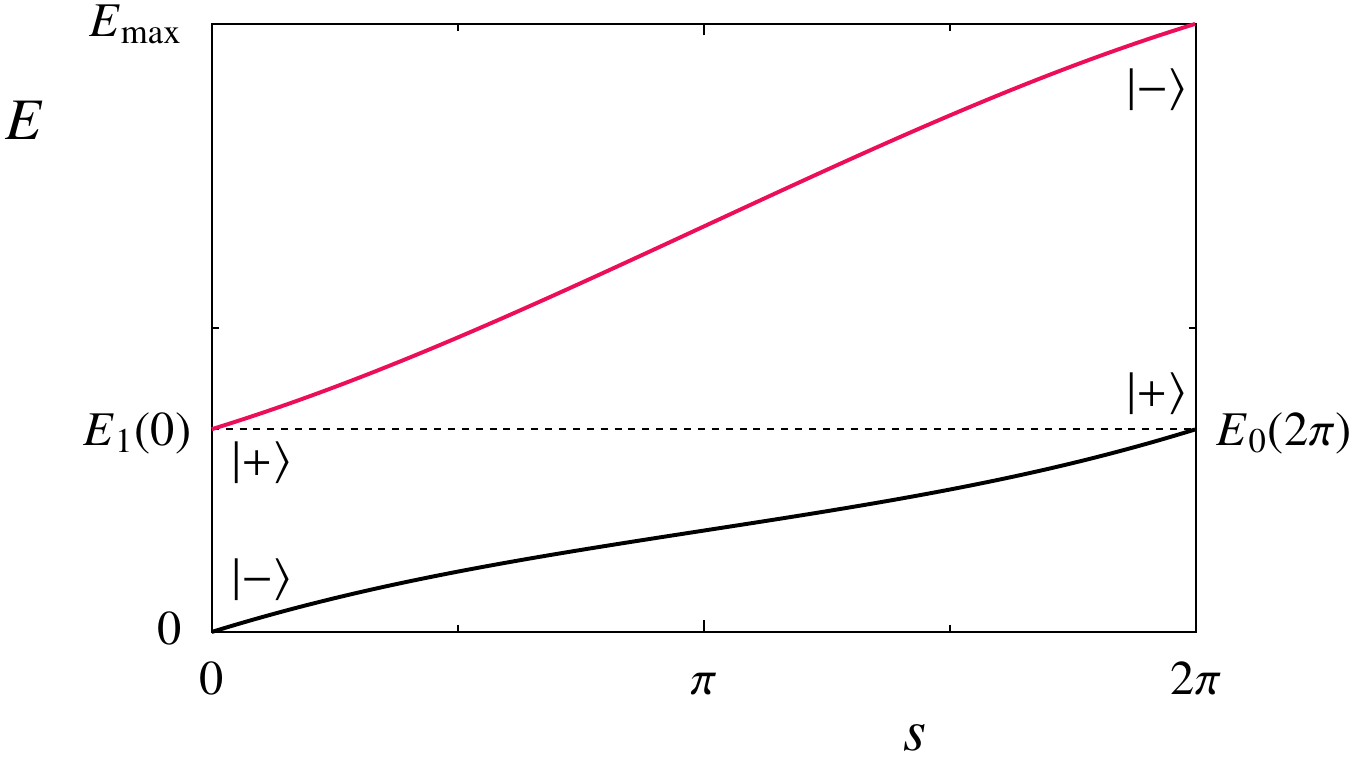}
  \caption{(Color online) 
    Two quasienergies $E_{\pm}(s)$ (bold curves) for the optimal choice
    of $\ket{v}$ given at Eq.~\eqref{eq:impracticableV}, with 
    $T=1$ and $E_{\rm P} = 2\pi/3$.
    The dashed line corresponds to a unperturbed eigenenergy 
    $E_1(0) = E_{\rm P}$.
    The period of quasienergy space is $E_{\rm max} = 2\pi$. 
    The ``ground'' quasienergy for the present choice of the branch
    exhibits anholonomy, connecting
    $E_0(s = 0) = 0$ and $E_0(s = 2\pi) = E_{\rm P}$.
    Accordingly the adiabatic passage along the ground quasienergy
    transports $\ket{-}$ at $s=0$ to $\ket{+}$ at  $s=2\pi$.
    Note that the ``first excited'' quasienergy that connects
    $(s,E) = (0, E_{\rm P})$ and $(2\pi, E_{\rm max})$ 
    has crossings with other quasienergies (not depicted), which are
    $s$-independent and, whose eigenspaces are orthogonal to 
    $\ket{\pm}$~\cite{Miyamoto-PRA-76-042115}.
  }
  \label{fig:extreme}
\end{figure}

The eigenvalues are
\begin{equation}
  z_{\pm}(s) \equiv 
  \exp\left[-i\left(\frac{\EP T + s}{2}\pm \Theta_{\rm P}(s)\right)\right]
  ,
\end{equation}
where $\Theta_{\rm P}(s)\equiv\cos^{-1}
\left(\cos\frac{\EP T}{2}\cos\frac{s}{2}\right)$.
The corresponding quasienergies 
$E_{\pm}(s) = (\EP T + s)/2\pm \Theta_{\rm P}(s)$
are depicted in Fig.~\ref{fig:extreme}.

The extraordinary speed up by $\ket{v}$~\eqref{eq:impracticableV}
has to be compensated by the cost for the preparation of
the kick part of the Floquet operator
$e^{-i s\ketbra{v}{v}}$, as is mentioned in the previous section.
This means that we have to
pay the cost to prepare the unitary operator 
that makes $\ket{x}$ from a simple state, say, $\ket{0}^{\otimes n}$.
Namely, the effort to solve the given problem by the standard
quantum computation is required. 
In this sense, the present choice of $\ket{v}$ is not practically useful.
However, the present argument is applicable to
study the theoretical nature of AAQC, as is shown 
in Sec~\ref{sec:AAQCEquivalence}.

\subsection{A ``fair'' choice of $\ket{v}$ for Grover's problem}
\label{sec:smallgap}

We first explain the standard adiabatic 
algorithm~\cite{Farhi-quant-ph-0001106} 
for Grover's quantum search~\cite{Grover-PRL-79-325}
before we examine a fair choice of the state $\ket{v}$.
Let $N$ be the number of items in the unstructured database. 
The items will be labeled
with integers $0,\ldots,N-1$, and the label of the answer is
denoted by $x$, which is assumed to be unique.
The Hilbert space of the arithmetic register for the standard adiabatic 
computation is spanned by 
an orthonormal system $\set{\ket{m}}_{m=0}^{N-1}$.
The finial Hamiltonian $\hat{H}_{\rm P}$ is a cost function
of Grover's problem: 
\begin{equation}
  \label{eq:GroverCostHamiltonian}
  \hat{H}_{\rm P} = \alpha (1 -\ketbra{x}{x}),
\end{equation}
where we introduce an energy scale $\alpha$($>0$).
With an initialization of the arithmetic register to be in $\ket\Ground$, 
an adiabatic passage along the path of 
the ground energy of $\hat{H}(s)$, with an adiabatic change of $s$,
transposes the state of the arithmetic register to $\ket\Answer$.
In SAQC, the minimum gap
for the ground energy of $H(s)$ is $\mathcal{O}(N^{-\frac{1}{2}})$
for asymptotically large $N$~\cite{Farhi-quant-ph-0001106}.
This implies that the running time is $\mathcal{O}(N)$,
if $ds/dt$ is kept constant along the path.
Namely, no quantum speedup is available.
However, a time-dependent $ds/dt$ introduced by Roland and Cerf
provides a Grover-type improvement of running time $\mathcal{O}(\sqrt{N})$,
once the locations of narrow gaps in the parameter space could be 
identified~\cite{Roland-PRA-65-042308}.
In the latter sense, the standard adiabatic quantum search has
the same computational power as the original Grover's algorithm.
We will show
that the anholonomic adiabatic approach 
has the same performance as the standard adiabatic approach.

In the Grover's problem, we have no prior knowledge of $\ket{x}$
to choose $\ket{v}$.
Hence, we employ $\ket{\Fair}$, which need to have
non-zero overlap with $\ket{x}$, to introduce
a normalized $\ket{v}$:
\begin{equation}
  \label{eq:fairketv}
  \ket{v}\equiv
  a \ket\Ground\otimes\ket{\Init} + b\ket\Fair\otimes\ket{\Fin}
  = a\ket{-}  + b\ket\Fair\otimes\ket{\Fin},
\end{equation}
where we choose the phase of $\ket\Fair$ so that $b$ is positive.
Although the overlap $\bracket\Answer\Fair$ is required
to be non-zero, $|\bracket\Answer\Fair|$ become small as 
$N\to\infty$, due to the
lack of the prior knowledge of $\Answer$ to prepare $\ket\Fair$.
A typical state for $\ket{\Fair}$ may be a superposition of
all candidates of the answer, that is 
\begin{equation}
  \label{eq:typicalFair}
  \ket\Fair=\frac{1}{\sqrt{N}}\sum_{m=0}^{N-1}\ket{m}
  ,
\end{equation}
in this sense we consider the state $\ket{\Fair}$ as a fair choice
for $\ket{v}$. The preparation of  $\ket{\Fair}$ and hence $\ket{v}$
is trivial, and hence only polynomially many quantum gates are required
to prepare $\hat{U}_s$.
Hence
it is natural to assume 
$\bracket\Answer\Fair = \mathcal{O}(N^{-1/2})$ as $N\to\infty$. 
We introduce a small parameter
$\epsilon = \mathcal{O}(N^{-1/2})$ to parameterize 
the overlap between $\ket\Answer$ and $\ket\Fair$ as
$\bracket\Answer\Fair = e^{i\theta}\sin\epsilon$, where $0 < \epsilon < \pi/2$ 
and $\theta = \arg \bracket\Answer\Fair$.

To study the spectrum of $\hat{U}_s$,
it is convenient to introduce a normalized vector 
$\ket{\Answer^{\perp}}\in\HilA$:
\begin{equation}
  \ket{\Answer^{\perp}} \equiv \frac{1}{\cos\epsilon}
  \left(1 - \ketbra{x}{x}\right)\ket\Fair.
\end{equation}
Hence we have an expansion of $\ket{v}$~\eqref{eq:fairketv}
by orthonormal vectors:
\begin{equation}
 \ket{v} = a\ket{-} + be^{i\theta}\sin\epsilon\ket{+} + 
 b\cos\epsilon\ket{f},
\end{equation}
where a normalized vector 
\begin{equation}
  \label{eq:defKetf}
  \ket{f}\equiv \ket{\Answer^{\perp}}\otimes\ket\Fin
  \in \HilA\otimes\HilC
\end{equation}
satisfies $\bracket{\pm}{f} = 0$.

The subspace spanned by $\ket{\pm}$ and $\ket{f}$ is invariant
under 
$\hat{U}_{s} = e^{-i\hat{H}_0 T} e^{-is\ketbra{v}{v}}$~\eqref{eq:rank1Floquet} 
of Grover's problem. 
This is because $\ket{\Answer^{\perp}}$ is an eigenvector
of the cost Hamiltonian for Grover's problem,
$\hat{H}_{\rm P}$~\eqref{eq:GroverCostHamiltonian}
and $\ket{f}$ is accordingly an eigenvector of 
$\hat{H}_0$~\eqref{eq:unperturbedAAC}.

\begin{figure}
  \centering
  \includegraphics[width=8.6cm]{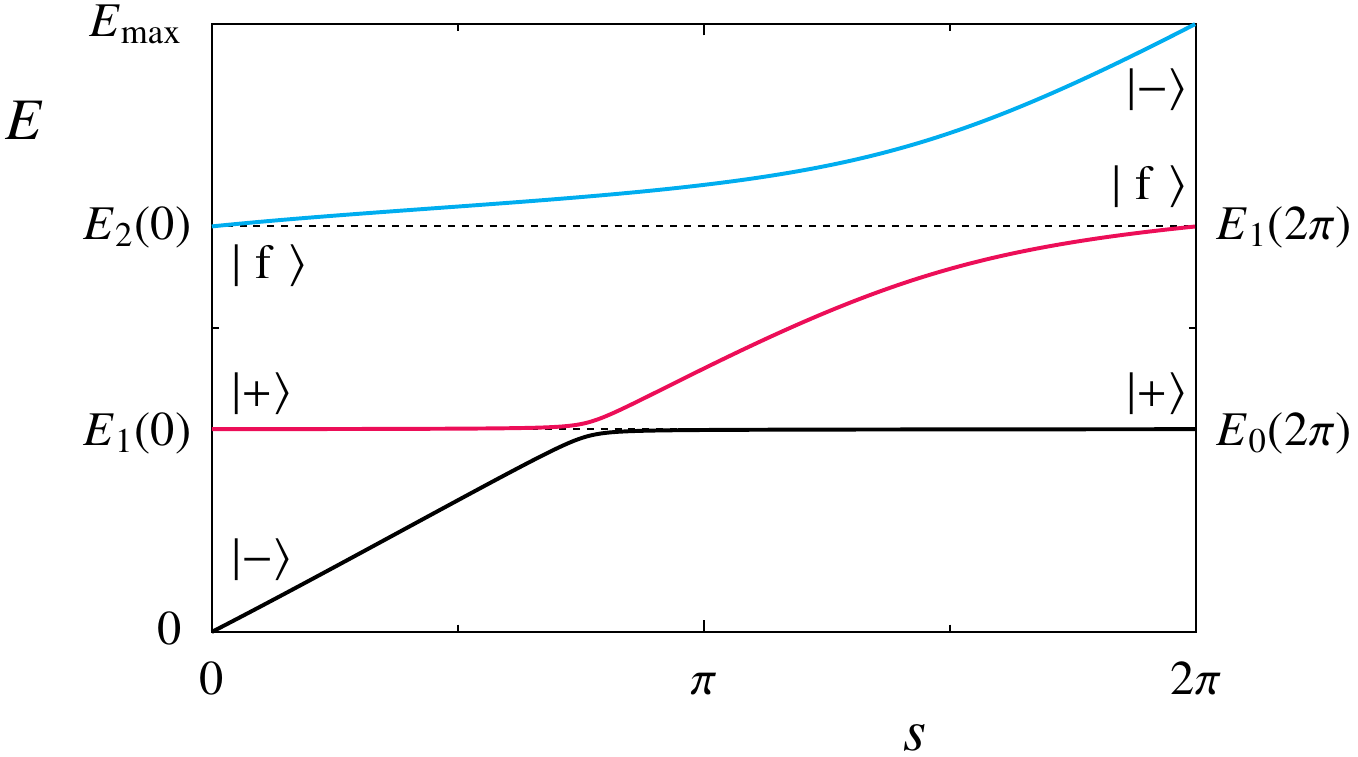}
  \caption{(Color online) 
    Three quasienergies $E_{\pm}(s)$ and $E_{\rm f}(s)$ (bold curves)
    for the fair choice
    of $\ket{v}$ (Eqs.~\eqref{eq:fairketv} and~\eqref{eq:typicalFair}), 
    with  $\left|\bracket\Answer\Fair\right|^2 = 1/100$ 
    (i.e., $N=100$) and 
    $\left|\bracket{-}{v}\right|^2 = 5/6$.
    The energy scale for the cost Hamiltonian~\eqref{eq:GroverCostHamiltonian}
    is $\alpha = 2\pi/3$.
    Other parameters are the same as in Fig~\ref{fig:extreme}.
    The adiabatic passage for the computation follows
    the ground quasienergy $E_0(s)$, which encounters 
    a narrow avoided crossing during the increment of
    $s$ from $0$ to $2\pi$.
  }
  \label{fig:fair}
\end{figure}

Hence it is suffice to examine a truncation of 
$\hat{U}_{0}$ by a three-level system
\begin{equation}
  \hat{U}_{0} = \ketbra{-}{-} + e^{-i\EP T}\ketbra{+}{+}
  + e^{-i(\EP+\alpha) T}\ketbra{f}{f}
\end{equation}
for Grover's problem.
The minimum quasienergy gap of 
$\hat{U}_{s}$ along an increment of $s$ is 
immediately obtained through numerical diagonalizations 
of $3\times3$ unitary matrices (Fig.~\ref{fig:fair}).

We show that
the adiabatic passage from
$\ket{-}$ to $\ket{+}$ encounters a narrow quasienergy gap 
in Fig.~\ref{fig:fair},
leaving the details of calculation to Appendix~\ref{app:gapForFairGrover}.
The
magnitude of the gap is $\mathcal{O}(\epsilon)=\mathcal{O}(N^{-1/2})$
by a perturbation expansion by $\epsilon$.
An application of  
Landau-Zener-St\"uckelberg formula for an estimation of nonadiabatic error
implies that 
the running time of adiabatic approach 
is $\mathcal{O}(N)$. However, if the location of the quasienergy
gap is identified, Roland and Cerf's prescription~\cite{Roland-PRA-65-042308}
is also applicable
to the quantum map $\hat{U}_{s}$ 
to obtain a Grover-type improvement of running time 
$\mathcal{O}(\sqrt{N})$.

\section{Efficient simulation of quantum circuits by AAQC}
\label{sec:AAQCEquivalence}

In this section, we will complete to show the equivalence of AAQC
with quantum circuits. 
Since we have already shown that
AAQC is efficiently simulated by quantum circuits, 
we only need 
to show the converse;
quantum circuits can be efficiently simulated by AAQC.
To achieve this, we utilize Aharonov {\it et al.}'s SAQC
~\cite{Aharonov-SJOC-37-166} to construct a simulator of
a given quantum circuit by AAQC. 
We remark that 
their proof for the equivalence of SAQC to quantum circuits can be 
directory applied to DAQC, since a subset of DAQC consists of
all discrete approximant of SAQC, i.e., SAQC is a subset of DAQC
in effect.
On the other hand, as for AAQC, the relationship with SAQC is not trivial.
This is the reason why the equivalence of AAQC to quantum circuits 
remains nontrivial.
The quasienergy gap of 
our simulator
is adjusted by $\ket{v}$, as is done in Section~\ref{sec:widegap},
with a reasonable cost for the present case.

Because the following argument involves extensive use of local observables,
it is convenient to denote them by local operators, instead of brackets.
Let $\Hq$ denote the Hilbert space of a qubit and
$\set{\ket{0},\ket{1}}$ corresponding unit vectors. 
Operators of the qubit are
\begin{gather}
  \hat{X}\equiv \ketbra{1}{0}+\text{h.c.},
  \quad
  \hat{Y}\equiv i\ketbra{1}{0}+\text{h.c.},
  \nonumber \\
  \hat{Z}\equiv \ketbra{0}{0}-\ketbra{1}{1},
  \nonumber \\
  \hat{\zero}\equiv \frac{1}{2}(1+\hat{Z}) = \ketbra{0}{0},
  \quad
  \hat{\one}\equiv \frac{1}{2}(1-\hat{Z}) = \ketbra{1}{1}.
\end{gather}
and we need to introduce ``annihilation'' and ``creation'' operators:
\begin{equation}
  \hat{A}\equiv 
  \ketbra{0}{1},
  \quad
  \hat{A}^{\dagger} \equiv
  \ketbra{1}{0}.
\end{equation}

The target of the simulation is a quantum circuit that involves
$n$ qubits, whose Hilbert space is denoted by $\HilA=(\Hq)^{\otimes n}$.
Let $\hat{a}_j^{({\rm A})}$ be 
a
local operator $\hat{a}$ on
$j$-th qubit.
The quantum circuit is composed by $L$ elementary circuits
$\set{\hat{U}_l}_{l=1}^L$, where $\hat{U}_l$ involves only few 
(possibly two) qubits, and its state evolves as
\begin{equation}
  \ket{\alpha(l)} = \hat{U}_l\ket{\alpha(l-1)}.
\end{equation}
where the initial and final states of the quantum circuit
are $\ket{\alpha(0)}\equiv\ket{0}^{\otimes n}$
and $\ket{\alpha} \equiv \ket{\alpha(L)}$, respectively.
Our simulator of the quantum circuit by an AAQC depends on
the initial and final Hamiltonians $\HB$ and $\HP$ of 
Aharonov {\it et al.}'s simulator of the same circuit 
by a SAQC. Hence the contsruction of these Hamiltonians explained.

Another ingredient of the simulator is Kitaev's clock~\cite{Kitaev-2002}, 
which is composed by $L$ qubits.
Let $\HilClock$ be the Hilbert space of the clock's states and
$\hat{a}_l$ be a local operator $\hat{a}$ on clock's $l$-th qubit.
The state of clock at step $l$ is associated with a state
\begin{equation}
  \ket{c(l)} \equiv \ket{1^l 0^{L-l}},
\end{equation}
which is an eigenstate of 
$\hat{\one}_1\hat{\one}_2\ldots\hat{\one}_l
\hat{\zero}_{l+1}\ldots\hat{\zero}_{L-1}\hat{\zero}_L$
with a non-degenerate eigenvalue $1$. Although $\set{\ket{c(l)}}_{l=0}^{L}$
is orthonormal, it is not a complete system of $\HilClock$.
Let us introduce a Hilbert space $\HilClock^{({\rm V})}$ that is spanned
by $\set{\ket{c(l)}}_{l=0}^{L}$. When the state of the clock does not
belong to $\HilClock^{({\rm V})}$, the clock is ``out of order''.
A cost Hamiltonian on $\HilClock$
\begin{align}
  \label{eq:DefHclock}
  \hat{H}_{\rm clock} \equiv
  \sum_{l=1}^{L-1} \hat{\zero}_l \hat{\one}_{l+1}
\end{align}
is introduced in order to characterize $\HilClock^{({\rm V})}$.
By definition,
the ground subspace of $\hat{H}_{\rm clock}$
is $\HilClock^{({\rm V})}$.

Combining Kitaev's clock and the quantum circuit, the
state of the
whole system ``at $l$-th step'' is 
$\ket{\gamma(l)}\equiv\ket{\alpha(l)}\otimes\ket{c(l)}$.
Let $\ket{\eta}$ be a superposition of $\ket{\gamma(l)}$'s:
\begin{align}
  \label{eq:DefEta}
  \ket{\eta}\equiv\frac{1}{\sqrt{L+1}}\sum_{l=0}^L \ket{\gamma(l)},
\end{align}
which is the destination state, i.e., the ground state of
$\HP$, of Aharonov {\it et al.}'s simulator of the quantum circuits.
Let $\Hilbert_0$ be $L$-dimensional Hilbert space spanned by 
$\set{\ket{\gamma(l)}}_{l=0}^L$. 
It will be clear that we need to focus on  $\Hilbert_0$ in
the following analysis of the simulator.

The initial state of Aharonov {\em et al.}'s simulator is 
$\ket{\gamma(0)}=\ket{0}^{\otimes (n+L)}$. 
To achieve this, 
$\hat{H}_{\rm B}$,
the initial Hamiltonian of Aharonov {\em et al.}'s SAQC,
is composed by three parts.
The first ingredient is $\hat{H}_{\rm clock}$~\eqref{eq:DefHclock},
whose ground state must be in $\HilA\otimes\HilClock^{({\rm V})}$.
The next is 
\begin{align}
  \hat{H}_{\rm clockinit}
  \equiv \hat{\one}_1,
\end{align}
whose ground subspace is $\HilA\otimes\ket{c(0)}$, if the state space is
restricted within $\HilA\otimes\HilClock^{({\rm V})}$.
To initialize the quantum circuit, we introduce
\begin{align}
  \hat{H}_{\rm input} \equiv
  \left(\sum_{j=1}^{n} \hat{\one}^{({\rm A})}_j \right)
  \otimes\hat{\zero}_1,
\end{align}
whose ground subspace contains
$\ket{\gamma(0)}$ and
$\HilA\otimes\left({\rm span}\set{\ket{c(l)}}_{l=1}^L\right)$.
To obtain $\HB$, three Hamiltonians are combined:
\begin{align}
  \label{eq:HBsim}
  \HB\equiv
  \hat{H}_{\rm clockinit} + \hat{H}_{\rm input} + \hat{H}_{\rm clock},
\end{align}
whose unique ground state is $\ket{\gamma(0)}$.

The finial Hamiltonian $\HP$ of Aharonov {\em et al.}'s SAQC is also 
composed by three parts. 
Two of them, $\hat{H}_{\rm input}$ and $\hat{H}_{\rm clock}$,
are also contained in $\HB$.
The last part, which we denote $\hat{H}_{\rm h}$,
of $\HP$ is determined so that 
$\ket\eta$ (Eq.~\eqref{eq:DefEta})
is a nondegenerate ground state of $\HP$.
A component of $\hat{H}_{\rm h}$, which correspond to 
the ``$l$-th step'' of the quantum circuit, is
a ``five-body'' interaction term 
($1 < l < L$):
\begin{align}
  \hat{H}_l&\equiv
  \frac{1}{2}\hat{\one}_{l-1}\hat{\zero}_l\hat{\zero}_{l+1} 
  - \frac{1}{2}
  \left(\hat{\one}_{l-1}\hat{U}_l\hat{A}^{\dagger}_l\hat{\zero}_{l+1} 
    + \text{h.c.}\right)
  \nonumber\\ &\quad{}
  + \frac{1}{2}\hat{\one}_{l-1}\hat{\one}_l\hat{\zero}_{l+1}
  \\ &
  = 
  \hat{\one}_{l-1}\hat{h}_l\hat{\zero}_{l+1},
\end{align}
where the ``three-body'' term
\begin{equation}
  \label{eq:AharonovThreeBody}
  \hat{h}_l\equiv
  \frac{1}{2} 
  \left[1 -\left(\hat{U}_l\hat{A}^{\dagger}_l + \text{h.c.}\right)\right]
\end{equation}
transfers $\ket{\gamma(l)}$ into a superposition of
$\ket{\gamma(l\pm 1)}$ and the projectors 
$\hat{\one}_{l-1}$ and $\hat{\zero}_{l+1}$ make
$\hat{H}_l$ to be invariant in $\Hilbert_0$.
With the boundary terms
\begin{align}
  \hat{H}_1 
  \equiv
  \hat{h}_1 \hat{\zero}_2,
  \quad\text{and}\quad
  \hat{H}_L 
  \equiv
  \hat{\one}_{L-1}\hat{h}_L,
\end{align}
we have 
\begin{equation}
  \hat{H}_{\rm h}\equiv \sum_{l=1}^L \hat{H}_l,
\end{equation}
which has a unique ground state $\ket\eta$ with the ground energy $0$ and
leaves $\Hilbert_0$ invariant.
Hence we obtain the final Hamiltonian
\begin{equation}
  \label{eq:defHP}
  \HP \equiv
  \hat{H}_{\rm h}
  + \hat{H}_{\rm input} + \hat{H}_{\rm clock}
  .
\end{equation}
In the following, we employ the restriction of $\HP$ to $\Hilbert_0$.

To incorporate Aharanov {\it et al.}'s $\HP$ with the AAQC simulator, 
we need to estimate
the asymptotic behaviors of 
its energy gap of the ground state and the maximum eigenenergy
in the limit $L\to\infty$, 
within the subspace $\Hilbert_0$.
The spectrum properties of $\HP$ 
are also crucial for the Aharonov {\it et al.}'s SAQC
and are already clarified by them~\cite{Aharonov-SJOC-37-166}.
Hence we explain their results only briefly.
Since the subspace $\Hilbert_0$ is an eigenspace of 
$\hat{H}_{\rm input} + \hat{H}_{\rm clock}$, which is
a part of $\HP$ (see, Eq.~\eqref{eq:defHP}), and 
the corresponding eigenvalue is zero, 
it is suffice to examine $\hat{H}_{\rm h}$, the nontrivial 
remainder of $\HP$.
It is straightforward to estimate the spectrum of $\hat{H}_{\rm h}$
from the fact that  $\hat{H}_{\rm h}$ is a discretized one-dimensional
Laplacian. The first excited energy of $\HP$ is 
\begin{equation}
  \label{eq:defHhGap}
  \Delta = \mathcal{O}(L^{-2})
\end{equation}
and is doubly degenerated.
The maximum eigenenergy of $\HP$ is
\begin{equation}
  \label{eq:defHhMax}
  W_{\rm P} = 2 +\mathcal{O}(L^{-1}).
\end{equation}
Hence the two Hamiltonians $\HB$ and $\HP$ for Aharonov {\em et al.}'s 
SAQC have introduced.

With these preparations, we introduce an AAQC, which simulates
the quantum circuit. 
The construction of the AAQC from Aharonov {\em et al.}'s SAQC
follows the prescription explained
in Section~\ref{sec:FundamentalsAAQC}.
The unperturbed Hamiltonian for the AAQC is
\begin{equation}
  \hat{H}_0 \equiv
  \HB \ctrlRegister{\hat\zero} + (\HP + E_{\rm P})\ctrlRegister{\hat\one}
\end{equation}
where $\ctrlRegister{\hat\zero}$ and 
$\ctrlRegister{\hat\one}$ are projectors of a control register made of 
a qubit, and $E_{\rm P}$ is non-negative but smaller than 
$1$, which is the first excited energy of $\HB$.
The ground and the first excited states of $\hat{H}_0$ are
\begin{equation}
  \label{eq:defMinusAAQC}
  \ket{-}\equiv\ket{\gamma(0)}\otimes\ket{0} = \ket{0}^{\otimes (n+L+1)}
\end{equation}
and
\begin{equation}
  \label{eq:defPlusAAQC}
  \ket{+}\equiv\ket\eta\otimes\ket{1},
\end{equation}
respectively, and they are non-degenerate.
Our unperturbed Floquet operator is
\begin{equation}
  \hat{U}_0 \equiv e^{-i\hat{H}_0 T},
\end{equation}
where $T$ is the period of our quantum map.
Since $\hat{U}_0$ is induced by $\hat{H}_0$, which is composed only 
by few-body interactions,
$\hat{U}_0$ can be implemented efficiently.
To ensure that $\hat{U}_0$ has no quasienergy between 
$0 < E < E_{\rm P}$, $T$ is assumed to be smaller than
$2\pi/W$, where $W$ is the maximum eigenenergy of $\hat{H}_0$.
From Eq.~\eqref{eq:defHhMax}, we have
$W =  E_{\rm P} + W_{\rm P} = E_{\rm P} + 2 + \mathcal{O}(L^{-1})$.

To build a quasienergy path that connects between 
$\ket{-}$ and $\ket{+}$ 
(Eqs.~\eqref{eq:defMinusAAQC} and \eqref{eq:defPlusAAQC}), we employ
\begin{align}
  \ket{v}=\frac{1}{\sqrt{2}}(\ket{-}+\ket{+})
\end{align}
to realize quasienergy anholonomy with $N$-independent gap in
the Floquet operator
\begin{equation}
  \hat{U}_{s} = 
  \hat{U}_0 e^{-is\ketbra{v}{v}}.
\end{equation}
It remains to show that the kick part $e^{-is\ketbra{v}{v}}$ is
efficiently implementable.
In the case of Grover's quantum search, this brings us 
disastrous slowdown
to prepare $\ket{v}$ (Section~\ref{sec:widegap}).
Contrary to this, we will show that the cost for the preparation of
$\ket{v}$ is negligibly small, in the sense of quantum circuit complexity,
to simulate the quantum circuit.
The crucial point is to obtain $\ket\eta$ (in $\ket{+}$) 
from $\ket{\gamma(0)}$.
This is decomposed into two steps. 
We first introduce 
a Fourier transformation on $\HilClock^{({\rm V})}$ 
as
\begin{align}
  \hat{F}_{\rm clock}\equiv
  \sum_{j=0}^{L} \ketbra{c_j}{c(j)}
  ,
\end{align}
where
\begin{align}
  \ket{c_k}\equiv
  \frac{1}{\sqrt{L+1}}\sum_{l=0}^{L}e^{-i2\pi kl/(L+1)}\ket{c(l)}
  .
\end{align}
The classical FFT algorithm ensures that 
the cost of $\hat{F}_{\rm clock}$ is $\mathcal{O}(L\ln L)$ steps.
$\hat{F}_{\rm clock}$ acts on $\ket{\gamma(0)}$, yielding
\begin{align}
  \hat{F}_{\rm clock}\ket{\gamma(0)} &
  = \frac{1}{\sqrt{L+1}}\ket{\alpha(0)}\otimes
  \sum_{l=0}^{L}\ket{c(l)}
  .
\end{align}
Second, we introduce a ``conditional evolution operator''
\begin{align}
  \hat{U}_{\rm h}&
  \equiv 
  \left(\hat{\zero}_L + \hat{\one}_L \hat{U}_L\right)
  \left(\hat{\zero}_{L-1} + \hat{\one}_{L-1} \hat{U}_{L-1}\right)
  \nonumber\\ &\qquad{}\times
  \cdots
  \left(\hat{\zero}_{1} + \hat{\one}_1 \hat{U}_{1}\right),
\end{align}
which is a product of $L$ few-body unitary operators and 
is accordingly efficiently implementable.
Because its 
action
on $\ket{c(l)}$ extracts a product of
quantum circuits as
$\hat{U}_{\rm h}\ket{c(l)}
= \hat{U}_{l}\hat{U}_{l-1}\cdots\hat{U}_{1}\ket{c(l)}$,
it turns out that an efficiently implementable unitary operator
$\hat{U}_{\rm h}\hat{F}_{\rm clock}$ gives us 
$\ket{\eta}$ from $\ket{\gamma(0)}$
\begin{align}
  \label{eq:generateHistory}
  \hat{U}_{\rm h}\hat{F}_{\rm clock}\ket{\gamma(0)} 
  = \frac{1}{\sqrt{L+1}}\sum_{l=0}^{L}\ket{\gamma(l)} 
  = \ket{\eta}.
\end{align}
Hence we have a unitary transformation $\hat{G}$, which 
makes $\ket{v}$ from $\ket{-}$, as follows:
\begin{align}
  \hat{G}= 
  \left(
    \ctrlRegister{\zero}
    + \hat{U}_{\rm h}\hat{F}_{\rm clock}\ctrlRegister{\one}
  \right)
  \ctrlRegister{\hat{H}}
\end{align}
where $\ctrlRegister{\hat{H}}$ is the Hadamard operation 
on the control register.
Now it is straightforward to see that $\hat{G}$ and
the kick operation 
\begin{align}
  e^{-is\ketbra{v}{v}}
  = \hat{G} e^{-is\ketbra{-}{-}}\hat{G}^{\dagger}
\end{align}
are also efficiently implementable.

Once we build the Floquet operator $\hat{U}_s$, 
$\ket{+}$ is obtained from $\ket{-}$, which is easy to
prepare, through an adiabatic
increment of $s$ from $0$ to $2\pi$ along the adiabatic passage
build on Cheon's quasienergy and eigenspace anholonomies of 
$\hat{U}_s$.
Now it is straightforward to obtain $\ket{\alpha}$, which the final state 
of the quantum circuit, from $\ket{+}$~\cite{note:amplifyAlpha}.

To determine the running time of our AAQC, the minimum quasienergy gap
along the adiabatic passage needs to be estimated.
Due to our choice of $\ket{v}$, the quasienergy spectrum of
$\hat{U}_s$ is simple. Two quasienergies depend on $s$. The other
quasienergies are independent with $s$, because the corresponding
eigenvectors have no overlap with $\ket{v}$~\cite{Miyamoto-PRA-76-042115}. 
Among them, the relevant quasienergy is
$E_{\rm P} +\Delta$, which correspond to the first excited
energy of $\HP$. The distance between $E_{\rm P} +\Delta$ and 
the quasienergy of the adiabatic passage takes minimum value
$\Delta$ at $s=2\pi$, at the end of the passage.
Hence the minimum gap is $\mathcal{O}(L^{-2})$.
Accordingly, the rate of adiabatic change of
$s$ needs to be algebraically slow in $L$ to ignore 
errors due to nonadiabatic transitions. 
Thus an efficient implementation of the simulator of the quantum circuit
is shown to be possible.

\section{Summary and outlook}
\label{sec:summary}

Adiabatic quantum computation that is originally composed of
a slowly varying Hamiltonian is reformulated purely in terms of discrete 
time evolution, e.g. concatenations of quantum circuits, 
where the concept of quasienergy, instead of eigenenergy, is employed 
to construct the adiabatic passage.
Furthermore, a nontrivial family of DAQC is introduced 
with the help of the quasienergy anholonomy, which is far more
easier to implement than the eigenvalue anholonomy in
adiabatic Hamiltonian evolutions.
We explain theoretical treatment of AAQC through 
estimations of quasienergy gaps for ``impractical'' and ``realistic''
cases. 
It turns out that the former case provides a key to show the equivalence
of AAQC and the conventional quantum computation. The proof of the equivalence
was
obtained 
by extending the proof by Aharonov {\it et al.} for SAQC.
Although our argument focus only on the essential point of 
Aharonov {\it et al.}'s proof, the rest of sophisticated points 
could be taken into account straightforwardly for AAQC.

There remains an open question whether AAQC really solves classically
intractable problems efficiently. 
To clarify this, we need to examine AAQC for a hard problem
by a classical computer. 
This certainly involves a simulation of complex quantum dynamics.
In this respect, we explain an advantage of AAQC over SAQC from a 
historical perspective obtained through the studies of classical 
and quantum chaos~\cite{Gutzwiller-1990}, 
i.e, prototypes of complex dynamics in classical and quantum mechanics. 
Although examples of chaos are described by differential 
equations~\cite{Lorenz-JAtmosSci-20-130},
the discovery of chaotic iterative 
mapping~\cite{Henon-CMP-50-69,Berry-AP-122-26} 
has galvanized the studies of chaos.
This approach allows us to facilitate both numerical and exact 
analysis~\cite{Tabor-1989}, which can also be applied to 
adiabatic quantum computation.

\begin{acknowledgments}
  AT wishes to thank Marko \v{Z}nidari\v{c} for discussion
  and valuable comments.
  This work is partly supported by JST.
\end{acknowledgments}

\appendix
\section{Estimation of quasienergies for Grover problem
  with ``fair'' $\ket{v}$}
\label{app:gapForFairGrover}

We will examine the quasienergies of 
AAQC's $\hat{U}_s$ (Eq.~\eqref{eq:rank1Floquet}) for 
Grover's problem with the fair choice of 
$\ket{v}$~(Eq.~(\ref{eq:fairketv})), introduced in Sec.~\ref{sec:smallgap}.
It is shown above that 
the subspace spanned by $\ket{\pm}$ and $\ket{f}$~\eqref{eq:defKetf}
is the relevant one for AAQC.

A perturbation expansion with a small parameter $\epsilon$ is suffice 
to obtain the asymptotic behavior of the minimum quasienergy gap for $N\gg 1$.
The reference of the expansion is a unitary operator 
\begin{equation}
  \hat{W}_{s}\equiv \hat{U}_0 e^{-is\ketbra{u}{u}}
\end{equation}
where $\ket{u} \equiv a\ket{-} + b\ket{f}$ agrees with $\ket{v}$
as $\epsilon\to 0$. 
Because of the fact $\bracket{u}{+}=0$, 
$\ket{+}$ is a trivial eigenvector of $\hat{W}_{s}$,
in the sense that the corresponding eigenvalue $e^{-i\EP T}$ 
is independent with $s$.
Let us denote
$\ket{w_-(s)}$ and $\ket{w_{f}(s)}$ be
the two eigenvectors of $\hat{W}_{s}$
in the remaining subspace spanned by $\ket{-}$ and $\ket{f}$, 
where we impose $\ket{w_-(0)}=\ket{-}$ and $\ket{w_{f}(0)} = \ket{f}$.
The corresponding quasienergies $W_-(s)$ and 
$W_{f}(s)$ are monotonically increase with $s$
and exhibit quasienergy anholonomy:
$W_-(0) = 0\le W_-(s)\le W_-(2\pi) = \EP + \alpha
= W_{f}(0)\le W_{f}(s) \le W_{f}(2\pi) = 2\pi/T$.
At the same time, two quasienergies $W_-(s)$ and 
$e^{-i\EP T}$ of $\hat{W}_{s}$ crosses between 
$0 < s = s_{\rm c} < 2\pi$. 
Hence the adiabatic passages along both 
$W_-(s)$ is inapplicable for the adiabatic search,
because the destination of the state vector $\ket{f}$ does not
provide $\ket{x}$ immediately. Hence we do need to incorporate
the effect of small $\epsilon$ to achieve the anholonomic adiabatic
quantum search with $\hat{U}_{s}$~\eqref{eq:rank1Floquet}.

We rearrange $\hat{U}_{s}$ to carry out the perturbation
expansion from $\hat{W}_{s}$:
\begin{equation}
  \hat{U}_{s} = \hat{W}_{s} \hat{S}_{s},
\end{equation}
where 
\begin{equation}
  \hat{S}_{s}\equiv 
  e^{+is\ketbra{u}{u}}e^{-is\ketbra{v}{v}}.
\end{equation}
The perturbation $\hat{S}_{s}$ makes an exact crossing of the quasienergies
$W_-(s)$ and $e^{-i\EP T}$ of $\hat{W}_{s}$,
at $s = s_{\rm c}$, to an avoided crossing 
of the exact quasienergies $E_-(s)$ and  $E_+(s)$ of
$\hat{U}_{s}$.
The resultant quasienergy $E_-(s)$ connects
$\ket{-}$ at $s=0$ and $\ket{+}$  at $s=2\pi$.
The narrowest quasienergy gap 
between $E_-(s)$ and $E_+(s)$ 
around $s = s_{\rm c}$ determines 
the speed of the anholonomic quantum search.
By using the fact that both $\ket{u}$ and $\ket{v}$ belongs to a 
two-dimensional subspace of $\HilA\otimes\HilC$, $\hat{S}_{s}$ is
expressed as $\hat{S}_{s} = e^{-i\hat{s}_{s}}$, where
$\hat{s}_{s}$ is the perturbation ``Hamiltonian''
\begin{align}
  &
  \hat{s}_{s}\times \left(\Theta / \sin\Theta\right)^{-1}
  \nonumber \\ &
  = \left(\ketbra{\delta}{u}+ \ketbra{u}{\delta}
    +\ketbra{\delta}{\delta}\right)\sin{}s
  \nonumber \\ &\quad
  +i(\ketbra{u}{\delta}-\ketbra{\delta}{u})
  (1 -\frac{1}{2}\bracket{\delta}{\delta})\left(1-\cos{}s\right)
  \nonumber \\ &\quad
  -\left(2\ketbra{u}{u}+\ketbra{u}{\delta}+\ketbra{\delta}{u}\right)
  \Im(\bracket{u}{\delta})\left(1-\cos{}s\right),
\end{align}
where $\ket{\delta}\equiv \ket{v}-\ket{u}$
and $\Theta$ satisfies
\begin{align}
  \cos\Theta &
  = 1 
  - 2 \left(\bracket{\delta}{\delta} - |\bracket{u}{\delta}|^2\right)\sin^2\frac{s}{2}.
\end{align}
From 
$\bracket{\delta}{\delta} = 4 b^2 \sin^2\frac{\epsilon}{2}$
and 
$\bracket{u}{\delta} = - 2 b^2 \sin^2\frac{\epsilon}{2}$,
we have
\begin{align}
  \cos\Theta &
  = 1 
  - 8 b^2\sin^2\frac{\epsilon}{2} 
  \left(1 - b^2\sin^2\frac{\epsilon}{2}\right)\sin^2\frac{s}{2}.
\end{align}
Hence, in the asymptotic regime $|\epsilon|\ll 1$, 
we obtain $\Theta / \sin\Theta = 1 + \mathcal{O}(\epsilon^2)$.
Finally, we obtain the leading contribution of $\hat{s}_{s}$
with respect to an expansion by $\epsilon$:
\begin{equation}
  \hat{s}_{s}
  = 
  2\epsilon b\sin\frac{s}{2}
  \left(\ketbra{+}{u}e^{i\theta-is/2} + \text{h.c.}\right)
  +\mathcal{O}(\epsilon^2).
\end{equation}
Thus we conclude that, if we avoid $s_c = 0, \pi$,
the minimum quasienergy gap between $E_-(s)$ and $E_+(s)$ 
is $|2\epsilon b\sin(s_c/2)| = \mathcal{O}(\epsilon)$.


\begin{thebibliography}{31}
\expandafter\ifx\csname natexlab\endcsname\relax\def\natexlab#1{#1}\fi
\expandafter\ifx\csname bibnamefont\endcsname\relax
  \def\bibnamefont#1{#1}\fi
\expandafter\ifx\csname bibfnamefont\endcsname\relax
  \def\bibfnamefont#1{#1}\fi
\expandafter\ifx\csname citenamefont\endcsname\relax
  \def\citenamefont#1{#1}\fi
\expandafter\ifx\csname url\endcsname\relax
  \def\url#1{\texttt{#1}}\fi
\expandafter\ifx\csname urlprefix\endcsname\relax\def\urlprefix{URL }\fi
\providecommand{\bibinfo}[2]{#2}
\providecommand{\eprint}[2][]{\url{#2}}

\bibitem[{\citenamefont{Messiah}(1999)}]{AdiabaticPassage}
\bibinfo{author}{\bibfnamefont{A.}~\bibnamefont{Messiah}},
  \emph{\bibinfo{title}{Quantum Mechanics}} (\bibinfo{publisher}{Dover},
  \bibinfo{address}{New York}, \bibinfo{year}{1999}),
  chap.~\bibinfo{chapter}{17}.

\bibitem[{\citenamefont{Farhi et~al.}(2000)\citenamefont{Farhi, Goldstone,
  Gutmann, and Sipser}}]{Farhi-quant-ph-0001106}
\bibinfo{author}{\bibfnamefont{E.}~\bibnamefont{Farhi}},
  \bibinfo{author}{\bibfnamefont{J.}~\bibnamefont{Goldstone}},
  \bibinfo{author}{\bibfnamefont{S.}~\bibnamefont{Gutmann}}, \bibnamefont{and}
  \bibinfo{author}{\bibfnamefont{M.}~\bibnamefont{Sipser}}
  (\bibinfo{year}{2000}), \eprint{arXiv:quant-ph/0001106}.

\bibitem[{\citenamefont{Aharonov et~al.}(2007)\citenamefont{Aharonov, van Dam,
  Kempe, Landau, and Lloyd}}]{Aharonov-SJOC-37-166}
\bibinfo{author}{\bibfnamefont{D.}~\bibnamefont{Aharonov}},
  \bibinfo{author}{\bibfnamefont{W.}~\bibnamefont{van Dam}},
  \bibinfo{author}{\bibfnamefont{J.}~\bibnamefont{Kempe}},
  \bibinfo{author}{\bibfnamefont{Z.}~\bibnamefont{Landau}}, \bibnamefont{and}
  \bibinfo{author}{\bibfnamefont{S.}~\bibnamefont{Lloyd}},
  \bibinfo{journal}{SIAM Journal on Computing} \textbf{\bibinfo{volume}{37}},
  \bibinfo{pages}{166} (\bibinfo{year}{2007}).

\bibitem[{\citenamefont{Farhi et~al.}(2001)\citenamefont{Farhi, Goldstone,
  Gutmann, Lapan, Lundgren, and Preda}}]{Farhi-Science-292-472}
\bibinfo{author}{\bibfnamefont{E.}~\bibnamefont{Farhi}},
  \bibinfo{author}{\bibfnamefont{J.}~\bibnamefont{Goldstone}},
  \bibinfo{author}{\bibfnamefont{S.}~\bibnamefont{Gutmann}},
  \bibinfo{author}{\bibfnamefont{J.}~\bibnamefont{Lapan}},
  \bibinfo{author}{\bibfnamefont{A.}~\bibnamefont{Lundgren}}, \bibnamefont{and}
  \bibinfo{author}{\bibfnamefont{D.}~\bibnamefont{Preda}},
  \bibinfo{journal}{Science} \textbf{\bibinfo{volume}{292}},
  \bibinfo{pages}{472} (\bibinfo{year}{2001}).

\bibitem[{\citenamefont{Sch{\"u}tzhold and
  Schaller}(2006)}]{Schuetzhold-PRA-74-060304}
\bibinfo{author}{\bibfnamefont{R.}~\bibnamefont{Sch{\"u}tzhold}}
  \bibnamefont{and} \bibinfo{author}{\bibfnamefont{G.}~\bibnamefont{Schaller}},
  \bibinfo{journal}{Phys. Rev. A} \textbf{\bibinfo{volume}{74}},
  \bibinfo{pages}{060304} (\bibinfo{year}{2006}).

\bibitem[{\citenamefont{\v{Z}nidari\v{c}}(2005)}]{Znidaric-PRA-71-062305}
\bibinfo{author}{\bibfnamefont{M.}~\bibnamefont{\v{Z}nidari\v{c}}},
  \bibinfo{journal}{Phys. Rev. A} \textbf{\bibinfo{volume}{71}},
  \bibinfo{pages}{062305} (\bibinfo{year}{2005}).

\bibitem[{\citenamefont{\v{Z}nidari\v{c} and
  Horvat}(2005)}]{Znidaric-PRA-73-022329}
\bibinfo{author}{\bibfnamefont{M.}~\bibnamefont{\v{Z}nidari\v{c}}}
  \bibnamefont{and} \bibinfo{author}{\bibfnamefont{M.}~\bibnamefont{Horvat}},
  \bibinfo{journal}{Phys. Rev. A} \textbf{\bibinfo{volume}{73}},
  \bibinfo{pages}{022329} (\bibinfo{year}{2005}).

\bibitem[{\citenamefont{Farhi et~al.}(2008)\citenamefont{Farhi, Goldstone,
  Gutmann, and Nagaj}}]{Farhi-IJQI-6-503}
\bibinfo{author}{\bibfnamefont{E.}~\bibnamefont{Farhi}},
  \bibinfo{author}{\bibfnamefont{J.}~\bibnamefont{Goldstone}},
  \bibinfo{author}{\bibfnamefont{S.}~\bibnamefont{Gutmann}}, \bibnamefont{and}
  \bibinfo{author}{\bibfnamefont{D.}~\bibnamefont{Nagaj}},
  \bibinfo{journal}{Int. J. Quantum Info.} \textbf{\bibinfo{volume}{6}},
  \bibinfo{pages}{503} (\bibinfo{year}{2008}), \eprint{arXiv:quant-ph/0512159}.

\bibitem[{\citenamefont{Hogg}(2003)}]{Hogg-PRA-67-022314}
\bibinfo{author}{\bibfnamefont{T.}~\bibnamefont{Hogg}}, \bibinfo{journal}{Phys.
  Rev. A} \textbf{\bibinfo{volume}{67}}, \bibinfo{pages}{022314}
  (\bibinfo{year}{2003}).

\bibitem[{not({\natexlab{a}})}]{note:discretizeSAQC}
\bibinfo{note}{We do not claim that any realization of SAQC requires the
  discretization. See, e.g., a proposal of an implementation with
  STIRAP~[D.~Daems and S.~Gu{\'e}rin, {Phys. Rev. Lett.} \textbf{99}, 170503
  (2007)].}

\bibitem[{\citenamefont{Steffen et~al.}(2003)\citenamefont{Steffen, van Dam,
  Hogg, Breyta, and Chuang}}]{Steffen-PRL-90-067903}
\bibinfo{author}{\bibfnamefont{M.}~\bibnamefont{Steffen}},
  \bibinfo{author}{\bibfnamefont{W.}~\bibnamefont{van Dam}},
  \bibinfo{author}{\bibfnamefont{T.}~\bibnamefont{Hogg}},
  \bibinfo{author}{\bibfnamefont{G.}~\bibnamefont{Breyta}}, \bibnamefont{and}
  \bibinfo{author}{\bibfnamefont{I.}~\bibnamefont{Chuang}},
  \bibinfo{journal}{Phys. Rev. Lett.} \textbf{\bibinfo{volume}{90}},
  \bibinfo{pages}{067903} (\bibinfo{year}{2003}).

\bibitem[{\citenamefont{Zel{'}dovich}(1967)}]{Zeldovich-JETP-24-1006}
\bibinfo{author}{\bibfnamefont{Y.~B.} \bibnamefont{Zel{'}dovich}},
  \bibinfo{journal}{Sov. Phys.--JETP} \textbf{\bibinfo{volume}{24}},
  \bibinfo{pages}{1006} (\bibinfo{year}{1967}).

\bibitem[{\citenamefont{Tanaka and Miyamoto}(2007)}]{Tanaka-PRL-98-160407}
\bibinfo{author}{\bibfnamefont{A.}~\bibnamefont{Tanaka}} \bibnamefont{and}
  \bibinfo{author}{\bibfnamefont{M.}~\bibnamefont{Miyamoto}},
  \bibinfo{journal}{Phys. Rev. Lett.} \textbf{\bibinfo{volume}{98}},
  \bibinfo{pages}{160407} (\bibinfo{year}{2007}).

\bibitem[{\citenamefont{Cheon and Tanaka}(2009)}]{Cheon-EPL-85-20001}
\bibinfo{author}{\bibfnamefont{T.}~\bibnamefont{Cheon}} \bibnamefont{and}
  \bibinfo{author}{\bibfnamefont{A.}~\bibnamefont{Tanaka}},
  \bibinfo{journal}{Europhys. Lett.} \textbf{\bibinfo{volume}{85}},
  \bibinfo{pages}{20001} (\bibinfo{year}{2009}).

\bibitem[{\citenamefont{Cheon}(1998)}]{Cheon-PLA-248-285}
\bibinfo{author}{\bibfnamefont{T.}~\bibnamefont{Cheon}},
  \bibinfo{journal}{Physics Letters A} \textbf{\bibinfo{volume}{248}},
  \bibinfo{pages}{285} (\bibinfo{year}{1998}).

\bibitem[{\citenamefont{Miyamoto and Tanaka}(2007)}]{Miyamoto-PRA-76-042115}
\bibinfo{author}{\bibfnamefont{M.}~\bibnamefont{Miyamoto}} \bibnamefont{and}
  \bibinfo{author}{\bibfnamefont{A.}~\bibnamefont{Tanaka}},
  \bibinfo{journal}{Phys. Rev. A} \textbf{\bibinfo{volume}{76}},
  \bibinfo{pages}{042115} (\bibinfo{year}{2007}).

\bibitem[{\citenamefont{Grover}(1997)}]{Grover-PRL-79-325}
\bibinfo{author}{\bibfnamefont{L.~K.} \bibnamefont{Grover}},
  \bibinfo{journal}{Phys. Rev. Lett.} \textbf{\bibinfo{volume}{79}},
  \bibinfo{pages}{325} (\bibinfo{year}{1997}).

\bibitem[{\citenamefont{Berry et~al.}(1979)\citenamefont{Berry, Balazs, Tabor,
  and Voros}}]{Berry-AP-122-26}
\bibinfo{author}{\bibfnamefont{M.}~\bibnamefont{Berry}},
  \bibinfo{author}{\bibfnamefont{N.}~\bibnamefont{Balazs}},
  \bibinfo{author}{\bibfnamefont{M.}~\bibnamefont{Tabor}}, \bibnamefont{and}
  \bibinfo{author}{\bibfnamefont{A.}~\bibnamefont{Voros}},
  \bibinfo{journal}{Ann. Phys.} \textbf{\bibinfo{volume}{122}},
  \bibinfo{pages}{26} (\bibinfo{year}{1979}).

\bibitem[{\citenamefont{Lloyd}(1996)}]{Lloyd-Science-273-1073}
\bibinfo{author}{\bibfnamefont{S.}~\bibnamefont{Lloyd}},
  \bibinfo{journal}{Science} \textbf{\bibinfo{volume}{273}},
  \bibinfo{pages}{1073} (\bibinfo{year}{1996}).

\bibitem[{not({\natexlab{b}})}]{note:eigenangle}
\bibinfo{note}{Our definition of eigenangle and the conventional one have
  opposite sign. This is to make a natural correspondence with quasienergy and
  the time development of dynamical phase.}

\bibitem[{\citenamefont{Nakamura}(2002)}]{Nonadiabatic}
\bibinfo{author}{\bibfnamefont{H.}~\bibnamefont{Nakamura}},
  \emph{\bibinfo{title}{Nonadiabatic transition}} (\bibinfo{publisher}{World
  Scientific}, \bibinfo{address}{Singapore}, \bibinfo{year}{2002}),
  \bibinfo{note}{and references therein.}

\bibitem[{\citenamefont{Breuer and Holthaus}(1989)}]{Breuer-ZPD-11-1}
\bibinfo{author}{\bibfnamefont{H.}~\bibnamefont{Breuer}} \bibnamefont{and}
  \bibinfo{author}{\bibfnamefont{M.}~\bibnamefont{Holthaus}},
  \bibinfo{journal}{Z. Phys. D} \textbf{\bibinfo{volume}{11}},
  \bibinfo{pages}{1} (\bibinfo{year}{1989}).

\bibitem[{not({\natexlab{c}})}]{note:twoquasienergy}
\bibinfo{note}{Note that $e^{-iE'T}$ and $e^{-iE''T}$ are also neighboring
  eigenenergy of $\hat{U}_0$ in the unit circle, due to the condition $T<
  2\pi/W$.}

\bibitem[{not({\natexlab{d}})}]{note:nanholonomy}
\bibinfo{note}{If $E_0(2\pi n)$ is the maximum eigenenergy of $\hat{H}_0$,
  $E_0(2\pi (n+1))$ is the ground energy $\hat{H}_0$ due to the periodicity of
  quasienergy space.}

\bibitem[{\citenamefont{Roland and Cerf}(2002)}]{Roland-PRA-65-042308}
\bibinfo{author}{\bibfnamefont{J.}~\bibnamefont{Roland}} \bibnamefont{and}
  \bibinfo{author}{\bibfnamefont{N.~J.} \bibnamefont{Cerf}},
  \bibinfo{journal}{Phys. Rev. A} \textbf{\bibinfo{volume}{65}},
  \bibinfo{pages}{042308} (\bibinfo{year}{2002}).

\bibitem[{\citenamefont{Kitaev et~al.}(2002)\citenamefont{Kitaev, Shen, and
  Vyalyi}}]{Kitaev-2002}
\bibinfo{author}{\bibfnamefont{A.~Y.} \bibnamefont{Kitaev}},
  \bibinfo{author}{\bibfnamefont{A.~H.} \bibnamefont{Shen}}, \bibnamefont{and}
  \bibinfo{author}{\bibfnamefont{M.~N.} \bibnamefont{Vyalyi}},
  \emph{\bibinfo{title}{Classical and Quantum Computation}}
  (\bibinfo{publisher}{American Mathematical Society},
  \bibinfo{address}{Providence, RI}, \bibinfo{year}{2002}).

\bibitem[{not({\natexlab{e}})}]{note:amplifyAlpha}
\bibinfo{note}{To obtain $\ket{\alpha}$ from a single measurement of $\ket{+}$,
  with reasonably small overhead and errors, it is suffice to employ Aharonov
  {\em et al.}'s amplification technique that increases the weight of
  $\ket{\alpha}$ in $\ket\eta$ (see Lemma 3.10 in
  Ref.~\cite{Aharonov-SJOC-37-166}).}

\bibitem[{\citenamefont{Gutzwiller}(1990)}]{Gutzwiller-1990}
\bibinfo{author}{\bibfnamefont{M.~C.} \bibnamefont{Gutzwiller}},
  \emph{\bibinfo{title}{Chaos in Classical and Quantum Mechanics}}
  (\bibinfo{publisher}{Springer-Verlag}, \bibinfo{address}{New York},
  \bibinfo{year}{1990}).

\bibitem[{\citenamefont{Lorenz}(1963)}]{Lorenz-JAtmosSci-20-130}
\bibinfo{author}{\bibfnamefont{E.~N.} \bibnamefont{Lorenz}},
  \bibinfo{journal}{J. Atmos. Sci.} \textbf{\bibinfo{volume}{20}},
  \bibinfo{pages}{130} (\bibinfo{year}{1963}).

\bibitem[{\citenamefont{H{\'e}non}(1976)}]{Henon-CMP-50-69}
\bibinfo{author}{\bibfnamefont{M.}~\bibnamefont{H{\'e}non}},
  \bibinfo{journal}{Comm. Math. Phys.} \textbf{\bibinfo{volume}{50}},
  \bibinfo{pages}{69} (\bibinfo{year}{1976}).

\bibitem[{\citenamefont{Tabor}(1989)}]{Tabor-1989}
\bibinfo{author}{\bibfnamefont{M.}~\bibnamefont{Tabor}},
  \emph{\bibinfo{title}{Chaos and integrability in nonlinear dynamics: an
  introduction}} (\bibinfo{publisher}{Wiley}, \bibinfo{address}{New York},
  \bibinfo{year}{1989}).

\end{thebibliography}



\end{document}